\documentstyle[aps,prd
,12 pt   
]{revtex}

\tightenlines

\newcommand{\beq}{\begin{equation}}
\newcommand{\eeq}{\end{equation}}

\def\lap{\lower.5ex\hbox{$\; \buildrel < \over \sim \;$}}
\def\gap{\lower.5ex\hbox{$\; \buildrel > \over \sim \;$}}
\def\O{{\cal O}}
\def\E{{\cal E}}

\def\O{{\cal O}}

\input epsf

\begin{document}

\title{Many worlds in one}

\author{Jaume Garriga$^{1,2}$ and Alexander Vilenkin$^{2}$}

\address{
$^1$ IFAE, Departament de F{\'\i}sica, Universitat Autonoma de Barcelona,\\
08193 Bellaterra (Barcelona), Spain\\
$^2$ Institute of Cosmology, Department of Physics and Astronomy,\\
Tufts University, Medford, MA 02155, USA
}

\maketitle

\begin{abstract}

A generic prediction of inflation is that the thermalized region
we inhabit is spatially infinite. Thus, it contains an infinite
number of regions of the same size as our observable universe,
which we shall denote as $\O$-regions. We argue that the number of possible
histories which may take place inside of an $\O$-region, from the
time of recombination up to the present time, is finite.
Hence, there are an infinite number of $\O$-regions with identical
histories up to the present, but which need not be identical in the future.
Moreover, all histories which are not forbidden by conservation laws will
occur in a finite fraction of all $\O$-regions.
The ensemble of $\O$-regions is reminiscent of the ensemble of universes
in the many-world picture of quantum mechanics. An important difference,
however, is that other $\O$-regions are unquestionably real.

\end{abstract}

\section{Introduction}

The presently observable spacetime region of the universe is defined
by the interior of our past light cone. It extends back to the time
of recombination, $t_{rec}$, beyond which the universe is opaque to
electromagnetic waves. For brevity, we shall denote as our $\O$-region
the interior of this past light cone from the time of recombination up
to the present. At time $t_{rec}$ this region was nearly
featureless with tiny deviations from homogeneity, presumably caused
by quantum fluctuations during inflation. It was practically
indistinguishable from same-size regions in other parts of the universe.

At later times, the small initial fluctuations were amplified by
gravitational instability, and the properties of $\O$-regions began to
diverge.  By the time of structure formation, the details of galaxy
distribution in different $\O$-regions varied considerably (although
statistically the regions were very similar).  Later on, the evolution
of life and intelligence could be influenced by statistical and
quantum fluctuations,
leading to further divergence of properties.  The histories of
different $\O$-regions are thus expected to be rather different.

It is conceivable, however, that there will be some regions with
identical histories.  Whether or not this situation is to be expected,
depends on the relative number of $\O$-regions and of the possible
histories.  In this paper, we are going to argue that the number of
distinct histories in an $\O$-region is finite, while the number of
$\O$-regions in the universe is infinite, and thus there should be an
infinite number of other regions with histories identical to ours.
Moreover, all histories which are not strictly forbidden by
conservation laws occur in a finite fraction of the $\O$-regions.
The argument below is far from being a rigorous proof of these
statements, but we do believe that it makes them rather plausible.

\section{The number of possible histories is finite}

It is relatively straightforward to argue that the total number of
``subjective'' histories which may take place in a given $\O$-region
is finite. For this purpose, we may consider a set of
"subjects" consisting of what Gell-Mann and Hartle \cite{GMH} have termed
IGUSes: information gathering and utilizing systems (humans are just particular
examples of IGUSes). The number of IGUSes which will develop in a $\O$
region is finite, since these IGUSes have to be built from finite resources.
If the ``minds'' of IGUSes can be modeled as discrete computers,
which operate much like our own brain by discrete firings of neurons,
then we may define the subjective history of the $\O$-region as the
sequence of all operations performed by all of the  minds
contained in the $\O$-region. Since the speed of computation is finite and
the available time is finite, the number of subjective histories will also be
finite.

In this Section, however, we shall argue for the stronger statement that
even the number of objective histories is finite.
If the universe obeyed the laws of classical physics, then the history
of an $\O$-region would be completely determined by the initial
conditions at $t_{rec}$.  In quantum mechanics, all histories that are
consistent with exact conservation laws can occur with some
probability, and in this sense the number of possibilities is greatly
increased.  On the other hand, classical description can
be made arbitrarily detailed, hence the number of
possible classical histories is infinite.  However, in quantum mechanics
there is a limit to the level of detail that can be achieved.

Modern formulations of quantum theory are
given in terms of alternative, decoherent, coarse-grained histories of
the system \cite{GMH,Griffiths,Omnes} (for a review and further
references, see \cite{GMHreview}).  To specify a
coarse-grained history, one can partition the spacetime region into cells
and divide the possible values of
physical quantities into finite bins.
A history is specified by indicating the appropriate bins
for the average values of all quantities in each of the cells.
A history is a semiclassical concept, and it has been argued
\cite{GMH,Jonathan} that the relevant quantities for a semiclassical
description are `hydrodymamic' variables, such as energy, momentum,
and various conserved and approximately conserved charges.
The histories
can be regarded as distinct if they decohere, that is, if the
decoherence functional $D(A,B)$ satisfies
$|D(A,B)|^2<\epsilon\ |D(A,A)D(B,B)|$ for
any two histories $A$
and $B$ in the set of all possible histories.  If this condition is
satisfied with a sufficiently small $\epsilon\ll 1$,
then a probability can be assigned to each history in the
set. The sizes of the cells and of the bins
for the values of the physical quantities cannot be made
arbitrarily small, since otherwise the property of decoherence would
be destroyed.  The best one can do is to obtain a maximally refined
set of histories, such that further refinement would make interference
between histories non-negligible.

For a system of a finite size observed for a finite interval of
time, the total number of cells is finite.  If the possible values of
all physical quantities are bounded, then the number of
bins for each quantity is finite and
it is clear that the total number of distinct histories is also
finite.

In the absence of gravity, the boundedness of physical quantities
usually follows from the finiteness of the energy of the system.
The gravitational energy is not positive-definite, and in the presence
of gravity the matter energy density can get arbitrarily large when it
is compensated by equally large negative gravitational energy
density.  This happens, for example, during gravitational collapse in
the interiors of black holes, in a recollapsing closed universe, and
could happen in some extremely large quantum fluctuations.  These
extreme situations, however, are far removed from the low-energy world
of the human experience, and it seems reasonable to isolate them when
comparing the histories of different ${\cal O}$-regions.  One could,
for example, assign all values of physical quantities corresponding to
over-Planckian energy density of matter to a single bin. Then,
histories that differ only in spacetime regions of higher than
Planckian density would be regarded as identical by humans (and
presumably by all other IGUSes).\footnote{It is also
possible that super-Planckian densities in cells need not be
considered because they are excluded by the fundamental theory.  If
cell sizes are large compared to the Planck length and the density is
greater than Planckian, $l\gg 1$ and $\rho>1$ in Planck units, then
the corresponding Schwarzschild radius $r_s\sim \rho l^3$ is much
greater than the cell size $l$.  One could expect that regions of this
kind would be hidden from external observers by event horizons.}  A
much lower threshold on the density would of course suffice, but the
important point for us here is that the relevant range of variation of
the physical quantities is finite.

The identification of two histories which differ only
in regions of energy density higher than a certain cut-off has been
justified by the fact that they are perceived as equal by IGUSes. This
is a somewhat "subjective" identification of histories, in the sense
described at the beginning of this section.
An alternative line of reasoning which is
completely objective, and which indicates that the number
of possible histories is finite, is based on the so-called holographic
bounds. The original bound, conjectured by 't Hooft \cite{tHooft} and Susskind
\cite{Susskind}, claims that the entropy in a finite region of space
satisfies
\beq
S\leq A/4,
\label{hb}
\eeq
where $A$ is the area (in Planck units) of a
surface currounding the region.  This bound is related to the entropy
bounds introduced by Bekenstein \cite{Be1Be2}.  As it stands, the
bound (\ref{hb}) cannot be true in general, since it is known to be
violated for collapsing regions inside their Schwarzschild radius and
for sufficiently large spherical regions of an expanding universe.
Bousso \cite{Bousso} suggested a modification of the bound which may
be free from these limitations.  The precise form of the bound will
not be important for our discussion here.  We shall simply assume that
the entropy in a finite region cannot be arbitrarily large,
\beq
S\leq \ln N,
\label{N}
\eeq
where $N$ may depend on the size of the region.  Since ${\cal
O}$-regions are finite, they should also obey the bound.

The bound (\ref{N}) implies that the number of linearly independent
quantum states for an ${\cal O}$-region does not exceed $N$.  In other
words, the dimension of the corresponding Hilbert space is $d\leq N$.
Thus, any complete set of projection operators
characterizing the configuration of the system at any given moment of time is
finite. A coarse-grained history can be specified by giving a complete set
of coarse-grained projection operators at discrete
intervals of time. Thus, if the number of time steps is finite, the number
of coarse-grained histories will also be finite.

Finally, let us estimate the number of possible histories
of an $\O$-region using the method of cells and bins.
If $N_c$ is the number of spacetime cells
and $N_b$ is the number of
relevant bins in field space, the number of coarse-grained histories
$N_h$ is given by
\begin{equation}
\ln N_{h} \sim N_{c} \ln N_{b}
\label{nh}
\end{equation}
The spacetime volume inside of our past light-cone from
the time of recombination is of order $t_0^4 \sim 10^{244} t_P$, where
$t_0 \sim 10^{10} yr$ is the age of the universe and $t_P \sim 10^{-44} s$ is
the Planck time. The number of spacetime cells of size $L$ is thus given
by
$$
N_{c} \sim 10^{244} (L M_P)^{-4}.
$$
To estimate the number of bins, consider for instance a scalar field
$\phi$. The equal time field commutator is
$
[\phi(x), {\dot\phi}(x')] = i \delta(x-x').
$
Integrating over $x$ and $x'$ over the volume $V = L^3$ and dividing over
$V^2$ gives
$$
[\Phi, {\dot\Phi}] = i /V,
$$
where $\Phi$ is the smeared field (averaged over the cell).  The uncertainty
relation following from the last equation is
$$
(\Delta\Phi) (\Delta{\dot\Phi})  \sim  1/V.
$$
If we try to specify $\Phi$ with an accuracy better than $1/L$,
then the uncertainty in $\dot\Phi$ is greater than $1/L^2$,
which means that the variation of $\Phi$ over the period $L$ is greater
than $1/L$. Thus,
a lower bound on the size of the bins which can be used
to specify a semiclassical history
is given by $1/L$, and
the number of bins will be bounded by
$$
N_{b} \lesssim \phi_{max} L,
$$
where $\phi_{max}$ is an estimate of the allowed range of
the field variable. The dependence of the number of bins on $L$
is only linear, and the number of histories $N_h$ will be maximized
for the lowest possible value of $L$ compatible with more than one
bin. In usual particle physics models $\phi_{max} \lesssim M_P$,
and in this case the number of coarse grained histories
for the scalar field is bounded by
\begin{equation}
\ln N_h \lesssim 10^{244}.
\label{huge}
\end{equation}
The range of the field could in principle be much larger than Planckian.
For instance, an exact Goldstone boson could have an infinite
range. In this case, the value of $\phi_{max}$ to be used above is still
finite. Requiring that the present average kinetic energy density be lower
than, say, the Planck energy density we have
$\phi_{max} \lesssim t_0 M_P^2 \sim 10^{61} M_P$.
When substituted in (\ref{nh}), and assuming that the minimum size of a cell
is the Planck lenght, this would increase the
exponent in (\ref{huge}) by a couple of units.

The standard model allows for many observables
other than canonical scalar fields. As mentioned above, it
is not clear how many of these observables should be used
to specify coarse-grained
histories and to calculate the number of bins. We may take the point of
view that
semiclassical coarse-grained histories are fully specified by the average
values of the electromagnetic field strength, the different baryonic and
leptonic charge densities and the gravitational field in
each cell \cite{GMH}. One
might also consider the local values of, say, the Higgs field and of the
non-abelian gauge bosons, but nothing essential seems to be lost
from the classical point of view if all possible values
of these fields are "traced over" in a single bin.
Therefore, the total number of relevant
observables seems to be very moderate. The size of the
cells which is necessary to ensure decoherence is likely to be much larger
than the Planck length, so the upper bound
(\ref{huge}) grossly overestimates the number of decoherent histories.
A more
realistic estimate may perhaps be obtained by considering cells of atomic
size $L\sim 10^{-8} cm$. This would cut the exponent in (\ref{huge}) down
to less than 150.

\section{The number of $\O$-regions is infinite}

We shall now argue, in the framework of inflationary cosmology, that
the number of ${\cal O}$-regions in the universe is infinite.

One of the striking aspects of inflation is that,
generically, it never ends (for a review of inflation, see
e.g. \cite{Alan}).  The evolution of the inflaton field
$\phi$ is influenced by quantum fluctuations, and as a result
thermalization does not occur simultaneously in different parts of the
universe.  On small scales, the fluctuations in the thermalization
time give rise to a spectrum of small density fluctuations, but on
large scales they make the universe extremely inhomogeneous.  In most
of the models, one finds that, at any time, there are parts of the
universe that are still inflating and that the total volume of the
inflating regions is growing with time \cite{AV83,Linde86}.
This picture is often
referred to as ``stochastic inflation'' or ``eternal inflation''.

\begin{figure}[t]
\centering
\hspace*{-4mm}
\leavevmode\epsfysize=10 cm \epsfbox{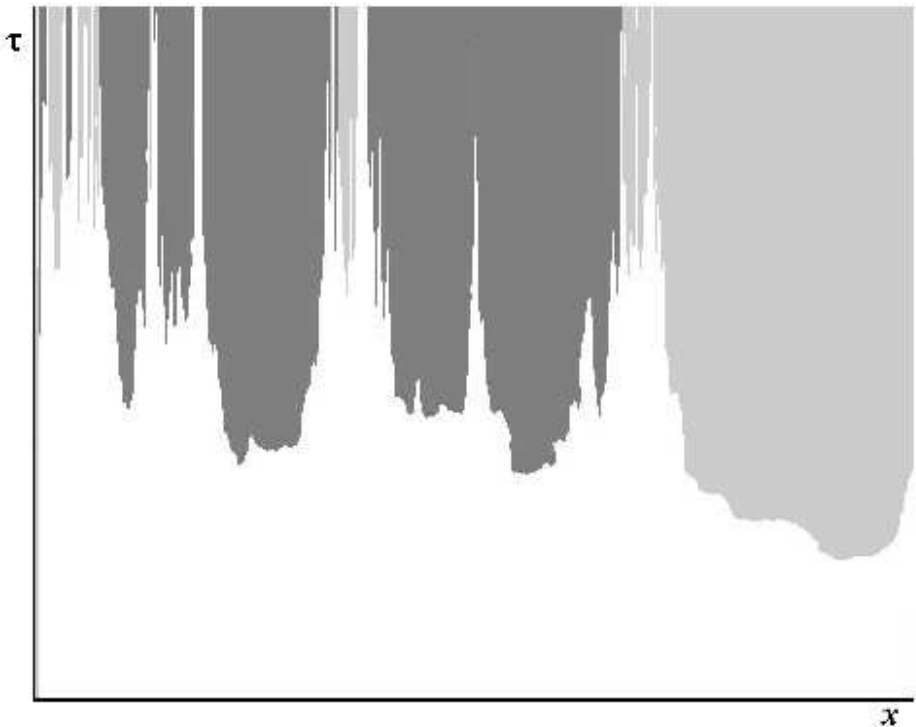}\\[3mm]
\caption[fig1]{\label{fig3} A numerical simulation of the spacetime
structure of an inflating universe \cite{vitaly}. The simulation
corresponds to a
double-well inflaton potential, with two degenerate minima where the
inflaton takes the values $\pm \eta$. Inflating regions
are white, while thermalized regions with inflaton values equal to
$+\eta$ and $-\eta$ are shown with different shades of grey .}
\end{figure}

The spacetime structure of an eternally inflating universe is
illustrated in Fig.1 \cite{vitaly}.
The vertical axis is time and the horizontal
axis is the comoving distance in one of the spatial directions.  The
boundaries between inflating and thermalized regions play the role of
the big bang for the corresponding thermalized regions.  In the
figure, these boundaries become nearly vertical at late times, so that
it appears that they become timelike.  The reason is that the
horizontal axis is the {\it comoving} distance, with the expansion of
the universe factored out.  The physical distance is obtained by
multiplying by the expansion factor $a(t)$, which grows exponentially
as we go up the time axis.  If we used the physical distance in the
figure, the thermalization boundaries would ``open up'' and become
nearly horizontal (but then it would be difficult to fit more than one
thermalized region in the figure).

The thermalization boundaries are surfaces of constant energy density of
matter.  It can be shown that they are infinite
spacelike surfaces and, with an appropriate choice of
coordinates, each thermalized region is an infinite sub-universe
containing an infinite number of galaxies.  These infinite
sub-universes can fit into finite comoving regions of the universe due
to the exponential expansion of the surrounding inflating regions.

\begin{figure}[t]
\centering
\hspace*{-4mm}
\leavevmode\epsfysize=10 cm \epsfbox{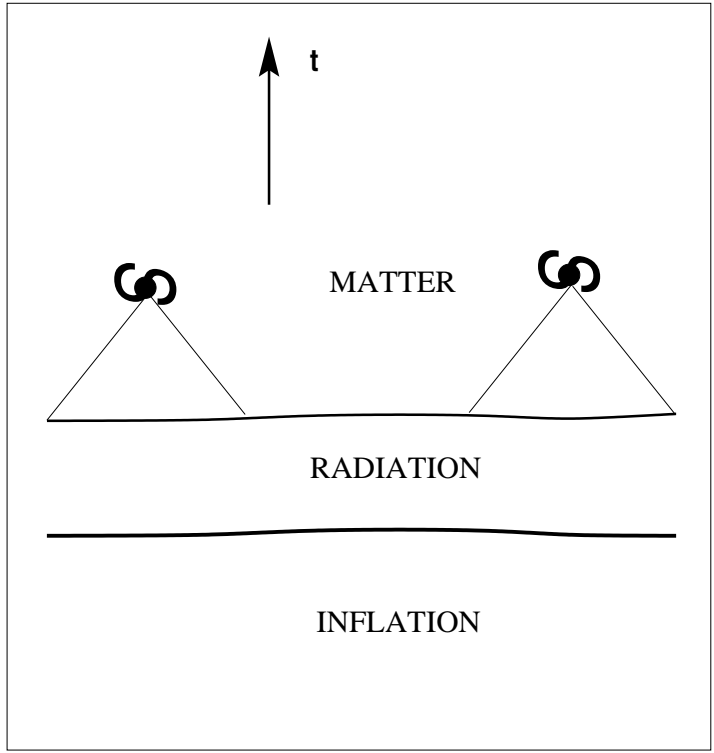}\\[3mm]
\caption[fig2]{\label{fig2}Spacetime structure near the thermalization
surface. After inflation, a period of radiation domination is
followed by the matter dominated era, where galaxies form
and civilizations flourish. Represented in the figure are the
$\O$-regions of two observers who live in mutually very distant galaxies.}
\end{figure}

The spacetime structure of a thermalized region near the
thermalization boundary is illustrated in Fig.2.  The thermalization
is followed by hot radiation era and then by a matter dominated era
during which luminous galaxies are formed and civilizations flourish.
All stars eventually die, and thermalized regions become dark, cold,
and probably not suitable for life.  Hence, observers are to be found
within a layer of finite (temporal) width along the thermalization
boundaries in Fig.1.  The tips of the light cones of the ${\cal
O}$-regions lie within that layer.
Since $\O$-regions have a finite size, it is easily seen that the
number of non-overlapping
$\O$-regions in each thermalized region is infinite.

Do histories unfolding in different ${\cal O}$-regions do so
simultaneously, or do they belong to different epochs?  The answer to
this depends on one's choice of the time coordinate.  The time
coordinate in Fig. 1 is the proper time along the worldlines of a
congruence of comoving observers.  With this choice, the constant time
surfaces contain inflating regions, as well as thermalized regions at
different stages of their evolution, from the high-energy state at
thermalization to extreme low densities and temperatures after the
death of stars.  A more natural choice in the interiors of thermalized
regions is to use the surfaces of constant energy density as constant
time surfaces.  (Such surfaces are Cauchy surfaces for the
corresponding thermalized regions, but not for the rest of the
universe.)  Then each thermalized region is described as an infinite
sub-universe, with all of its ${\cal O}$-regions evolving
simultaneously.

\section{Some implications}

We have argued that {\it (i)} the number of possible histories for an
${\cal O}$-region is finite and {\it (ii)} the number of $\O$-regions
in each of the thermalized regions of the universe is infinite.
With an appropriate choice of the time coordinate, all $\O$-regions in
a given thermalized region originate at the same time ($t=t_{rec}$)
and their histories are unfolding, so to say, as we speak.

The initial state of each $\O$-region can be characterized by a
density matrix, and the probabilities for all possible histories can
in principle be determined following the usual rules of quantum
mechanics.\footnote{We expect the histories in different $\O$-regions
to be statistically independent.  For nearby regions, there will be
some correlations in the low multipoles of the microwave sky, but such
correlations can be avoided by selecting an ensemble of well-separated
regions.}  All histories consistent with exact conservation laws will
have non-vanishing probabilities and will occur in an infinite number
of $\O$-regions.  It follows that there should be an infinite number
of $\O$-regions whose history is identical to ours.

This picture is reminiscent of the many-world interpretation of
quantum mechanics \cite{Everett,DeWitt,Deutsch1}.  According to this
interpretation, the wave function of the universe describes a
multitude of disconnected universes with all possible histories.  The
reality of the other universes is still a matter of controversy
(see, e.g., \cite{Ghost}), but the issue is rather academic, since the
existence of these universes cannot be tested in any direct way.

The ensemble of $\O$-regions that we discussed in this paper is very
similar to the ensemble of universes in the many-world picture.  An
important difference, however, is that other $\O$-regions are
unquestionably real.  There are infinitely many of them in our own
thermalized region.  Non-overlapping $\O$-regions are causally
disconnected, so at present we cannot get any information from other
$\O$-regions.  However, information exchange will become possible in the
future.  Given enough time, we might even be able to travel to what
used to be other $\O$-regions and to compare their histories with ours.
\footnote{We emphasize that the picture of the universe presented here is
independent of the interpretation of quantum mechanics.  If the many world
interpretation is adopted, then there is an ensemble of eternally
inflating universes, each having an infinite number of ${\cal O}$-regions.
Our picture should apply to each of the universes in the ensemble.}

The existence of $\O$-regions with all possible histories, some of
them identical or nearly identical to ours, has some potentially troubling
implications.  Whenever a thought crosses your mind that some terrible
calamity might have happened, you can be assured that it {\it has} happened
in some of the $\O$-regions.  If you nearly escaped an accident, then
you were not so lucky in some of the regions with the same prior
history.  On a positive side, some amusing
situations can be entertained where distant copies of ourselves
play all sorts of different roles.
Some readers will be pleased to know that there are infinitely many
$\O$-regions where Al Gore is President and - yes -
Elvis is still alive.

Consider the ensemble of $\O$-regions containing a civilization
whose history is identical to ours up to the present time (call this
ensemble $\E$).  In the
future the paths of these civilizations will of course diverge,
spanning all possible histories.  Each civilization in $\E$ can be
quantitatively characterized by a set of parameters such as its
lifetime, its age at the time when it made certain technological
advances, the extent to which it colonized the galaxy, etc.  One could
in principle determine the probability distribution for these parameters.
{\it A priori}, one expects that our civilization is a typical
representative of the ensemble, with most of the parameters near the
peak of the distribution.  This is an extension of the `principle of
mediocrity' \cite{AV95} (see also
\cite{Leslie,Gott})  to the realm of history.

We conclude with some comments on the future of civilizations in the
ensemble ${\cal E}$.  It appears very unlikely that a civilization can
survive forever \cite{Dyson,kraussstarkman}.  Even if it avoids natural
catastrophes and
self-destruction, it will in the end run out of energy.  The stars will
eventually die, other sources of energy (such as tidal forces) will also
come to an end, and our thermalized region will become cold, dark, and
inhospitable to life.  This argument, however, is statistical in nature,
since it is based on the second law of thermodynamics.  We know that
entropy can decrease due to thermal fluctuations, and it is in principle
possible that such fluctuations can sustain a civilization for an
arbitrarily long time.  The probability that a civilization will survive
on this kind of "fuel" for a time $t$ is of course a sharply decreasing
function of $t$.  However, for any finite $t$ the probability is finite,
and thus an infinite number of civilizations in our thermalized region
will live longer than any given time.  Needless to say, there is little
hope that we are going to be among the lucky civilizations whose life will
be prolonged by thermal fluctuations.

If we are doomed to perish, then perhaps we could perpetuate our legacy by
sending messages to new thermalized regions in our future?  In Ref.
\cite{future}
we explored some scenarios of this sort, where messages (tightly packed in
durable containers) could be sent into new inflating regions spawning
from our own thermalized region. A new inflating region within our
thermalized region can be
created by quantum tunneling or by virtue of strong violations of the weak
energy condition (see \cite{future} and references therein).
New civilizations arising after thermalization in the newly
created inflating regions might then be able to read those messages.
Recipients of our messages
could also send messages to the future, and so on. We would then
become a branch in an infinite ``tree'' of civilizations, and our
accumulated wisdom would not be completely lost.

However, in view of the discussion above,
it seems that sending messages into new
inflating regions is a rather useless activity.
If the history where our message is received in some $\O$-region is not
forbidden by the laws of Physics, then we know that it will be repeated
infinitely many times in different $\O$-regions of the same thermalization
surface. Hence, any finite number of sent messages does not
increase the chance of receiving a message \cite{Deutsch}.
An infinite number of
additional messages will come into existence by chance through thermal
or quantum fluctuations.
Whatever our motivation for sending the message could be, it is rather
pointless to do so because an infinite number of messages gets received anyway.
\footnote{Of course, not only the message representing our would-be legacy will
be spontaneously created.~ Nonsensical messages will also be created at a
much higher rate \cite{Deutsch} (perchance resembling - alas - copies of the
present paper. IGUSes beware!).}

\section*{Acknowledgements}

We are grateful to Ken Olum for interesting discussions and to
David Deutsch and Jim Hartle for useful correspondence.
This work was supported by the Templeton Foundation under grant COS 253.
J.G. is partially supported by CICYT, under grant AEN99-0766, and by the
Yamada Foundation. A.V. is partially supported by the National Science
Foundation.

\section*{Note Added}

After this paper was submitted, we learned about the early work by Ellis
and Brundrit \cite{eb} who discussed some of the issues we raised here
in the context of an open Robertson-Walker universe.  By assumption, such
a universe is infinite and homogeneous (on average), and thus contains an
infinite number of galaxies.  If our universe is of this kind, then Ellis
and Brundrit argued that there should be some regions in the universe with
histories very similar to that in our region.
Our analysis here goes beyond that of Ellis and Brundrit in two respects:
(i) the infinite number of ${\cal O}$-regions in our picture is a generic
consequence of inflation and does not have to be independently postulated,
and (ii) we argued that the number of distinct histories is finite, which
allowed us to conclude that there should be regions with histories {\it
identical} to ours.

\end{document}